\documentclass[a4paper, amsfonts, amssymb, amsmath, reprint, showkeys, nofootinbib, twoside]{revtex4-1}
\usepackage[english]{babel}
\usepackage[utf8]{inputenc}
\usepackage[colorinlistoftodos, color=green!40, prependcaption]{todonotes}
\usepackage{amsthm}
\usepackage{mathtools}
\usepackage{physics}
\usepackage{xcolor}
\usepackage{graphicx}
\usepackage[left=23mm,right=13mm,top=35mm,columnsep=15pt]{geometry} 
\usepackage{adjustbox}
\usepackage{placeins}
\usepackage[T1]{fontenc}
\usepackage{lipsum}
\usepackage{csquotes}

\usepackage{amssymb}
\usepackage{natbib}

\usepackage[LGRgreek]{mathastext} 
\newcommand\tab[1][1cm]{\hspace*{#1}} 

\usepackage{lineno}
\usepackage{setspace}
 \usepackage{hyperref}

\begin{document}
\title{Photocatalytic water splitting ability of  Fe/MgO-rGO nanocomposites towards hydrogen evolution}

\author{Fahmida Sharmin, Dayal Chandra Roy and M. A. Basith}
    \email[Email address: ]{mabasith@phy.buet.ac.bd}
    \affiliation{Nanotechnology Research Laboratory, Department of Physics, Bangladesh University of Engineering and Technology, Dhaka-1000, Bangladesh.\\ \\ DOI: \href{https://doi.org/10.1016/j.ijhydene.2021.09.072}{10.1016/j.ijhydene.2021.09.072}}


\begin{abstract}
Photocatalytic water splitting has greatly stimulated as an ideal technique for producing hydrogen (H\textsubscript{2}) fuel by employing two renewable sources, i.e., water and solar energy. Here, we have adopted a facile hydrothermal approach for the successful synthesis of reduced graphene oxide (rGO) incorporated Fe/MgO nanocomposites followed by thermal treatment at inert atmosphere to investigate their ability for photodegradation and photocatalytic hydrogen evolution via water splitting. Transmission Electron Microscopy images of Fe/MgO-rGO nanocomposite ensured the distribution of Fe/MgO nanoparticles throughout rGO sheets. Notably, all rGO supported nanocomposites, especially the one, thermally treated at 500 $^{\circ}$C at Argon (Ar) atmosphere has demonstrated significantly higher photocatalytic efficiency towards the photodegradation of a toxic textile dye, rhodamine B, than pristine MgO and commercially available Degussa P25 titania nanoparticles as well as other composites. Under solar irradiation, Fe/MgO-rGO(500) nanocomposite exhibited 86\%   degradation of rhodamine B dye and generated almost four times higher H\textsubscript{2} via photocatalytic water splitting compared to commercially available P25 titania nanoparticles. This promising photocatalytic ability of the Fe/MgO-rGO(500) nanocomposite can be attributed to the improved morphological and surface features due to heat treatment at inert atmosphere  as well as escalated charge carrier separation with increased light absorption capacity imputed to rGO incorporation.
\end{abstract}


\maketitle

\section{Introduction}
With the industrial revolution and an increase in the world population, the consistent growth in energy consumption, coupled with high pollution levels, has sparked tremendous research interest in alternative and renewable energy fuels \cite{Jain,Tama,Tong,Abe}. Solar energy is considered as a potential alternative energy source due to its abundance and sustainability. The efficient conversion of solar energy into chemical fuel is the ultimate challenge in confronting the energy crisis of today’s world. Nowadays, the production of H\textsubscript{2} through photocatalytic water splitting \cite{Fu,Sub} using  solar energy is gaining sincere attention in the scientific community because they are simple, efficient and have significant potential for future development \cite{Ma, Lv, Nayak,Winter}. Owing to its low cost and environmentally friendly nature, hydrogen has emerged as the most feasible substitute for traditional fossil fuels as a green and efficient energy source with high storage capacity \cite{Basith2,Shaheer, Jain1,Bicakova,Mori}. Though at present, renewable energy contributes only $\sim$5\% of commercial hydrogen production mainly through water electrolysis \cite{Cao2},  the remaining $\sim$95\% hydrogen is primarily generated from fossil fuels \cite{Dawood} which is neither cost-effective nor eco-friendly. Therefore, the development of efficient and cost-effective photocatalysts using renewable sources remains challenging.
On the other hand, photodegradation using photocatalysts for environmental remediation has been recognized as an effective approach for converting hazardous industrial contaminants into non-toxic byproducts \cite{Reddy, Patil, Yang}. 
In this context, several quantified photocatalysts, including TiO\textsubscript{2}, ZnO, CdS, SrTiO\textsubscript{3}, CdS/TiO\textsubscript{2}/Pt, and TiO\textsubscript{2}/Ag\textsubscript{2}O, have been widely used \cite{PA,Ga,Lang,Saad,Sola,Mazierski}. Unfortunately, these existing photocatalysts suffer from limitations such as large band gap and quick recombination of $e^{-}$ - $h^{+}$ pairs, limiting their photocatalytic performance and solar energy utilization eﬃciency. \\
\tab Magnesium oxide (MgO) with high surface reactivity, extensive absorption capacity, and ease of production has been used recently for various applications, including catalysis, ceramics, waste-water remediation, anti-bacterial agents, and so on \cite{Mohamed, Mou, Sutradhar}. However, due to its large band gap energy and fast charge carrier recombination, bare MgO has negligibly lower photocatalytic performance \cite{Mag}. These issues were resolved by adding low concentrations of transition metal ions at the conducting MgO interface, which prevents the formation of recombination centers for photo-generated $e^{-}$ - $h^{+}$ pairs \cite{Pin}. Among transition metal ions, iron (Fe) has a high transition ability and can trap multiple types of electron-hole pairs \cite{Ba,Ot,Man}. Thus incorporating Fe on MgO may improve the separation of the maximum number of electrons and holes, thereby avoiding recombination \cite{Borhade}. In recent years, graphene, a two-dimensional (2D) network of hexagonally structured sp\textsubscript{2} -hybridized carbon atoms, has sparked significant research interest in various energy adaptation applications as well as in the field of photocatalysis \cite{ Ramirez,Xu, Marlinda}. The importance of graphene is attributed to its excellent charge carrier mobility, electrical conductivity, high specific surface area and superior chemical stability \cite{ Purwajanti, Go}. Graphene has two oxidative derivatives, between which reduced graphene oxide (rGO) demonstrates enhanced photocatalytic activity compared to graphene oxide (GO) when they are being tailored as  supporting material to make a composite with photocatalysts \cite{Wang2, Hafeez}. The oxygen components over a pure GO surface may give a reverse action against its high electron conductivity, turning GO into an insulator \cite {Sadhukhan, 1Yu}. Several attempts have been adopted to synthesize pure graphene by reducing oxygen epoxy; among them, thermal treatment at inert atmosphere has been considered to be quite promising for producing high strength reduced graphene oxide by eliminating oxygen-containing functional groups from the surface of GO \cite{Shi}. Besides, thermal treatments may significantly modify the properties of photocatalysts since they have a notable impact on the physicochemical characteristics of the respective materials \cite{Feng, Yeoh}. \\
\tab Motivated by manifestations mentioned above, in this investigation, a simple hydrothermal route has been adopted for the synthesis of rGO supported Fe/MgO nanocomposites followed by heat treatment at different temperatures at Argon (Ar) atmosphere. Several extensive studies have been carried out to synthesize nanoscale MgO powders using different synthesis techniques, such as solid state, sol-gel, chemical precipitation method, etc \cite{Hong,Karnaukhov,Kumar1}. Among various methods, the hydrothermal technique is regarded as one of the most successful approaches to obtain the desired structures with high phase purity and uniform size distribution with an additional advantage of being low cost \cite{Basith1}. According to our acknowledgement, here, for the first time, we report the application of hydrothermally synthesized Fe/MgO-rGO nanocomposite photocatalysts on wastewater remediation and generation of H\textsubscript{2} through water splitting. Furthermore, the influence of heat treatment on the structure, morphology as well as the catalytic properties of Fe/MgO-rGO nanocomposites has also been explored. The photocatalytic performance of widely used Degussa titania (P25) nanoparticles has also been studied under similar experimental conditions  for comparative analysis. It has been observed that Fe/MgO-rGO after heat treatment at 500 $^{\circ}$C  at Ar atmosphere demonstrated higher potential in H\textsubscript{2} production via solar driven water splitting compared to other synthesized nanocomposites and P25 nanoparticles. In most of the previous investigations \cite{Bri,Sol}, the photocatalytic performance of rGO based composites have been observed to be  improved either by addition of H\textsubscript{2}O\textsubscript{2} doses in dye and photocatalyst mixture, or  by enhancing the power of light irradiation and controlling the pH of the solution, etc. However, in this investigation, Fe/MgO-rGO(500) nanocomposite has demonstrated satisfactory performance in RhB degradation and H\textsubscript{2} evolution without using any additional reagents or external energy input and in neutral pH condition. We anticipate that this unique property might distinguish our synthesized composite from those reported by others. Finally, a plausible reaction mechanism for the profound photocatalytic activities of Fe/MgO-rGO(500) has been suggested.

\section{Materials and Methods}
\subsection{Materials}
The chemical reagents used were analytical grade Manganese nitrate hexahydrate [Mg(NO\textsubscript{3})\textsubscript{3}.6H\textsubscript{2}O], Ferric nitrate nonahydrate [Fe(NO\textsubscript{3})\textsubscript{3}.9H\textsubscript{2}O], Graphite flakes, Sodium nitrate [NaNO\textsubscript{3}], Sodium hydroxide [NaOH], Potassium permanganate [KMnO\textsubscript{4}], Sulfuric acid [H\textsubscript{2}SO\textsubscript{4}], Hydrochloric acid [HCl] and Hydrogen peroxide [H\textsubscript{2}O\textsubscript{2}]. All the reagents were manufactured by Sigma-Aldrich, Germany.

\subsection{Sample preparation}

\subsubsection{Synthesis of Fe/MgO nanocomposites}

Fe/MgO nanocomposites were synthesized by adopting a facile and low-cost hydrothermal technique \cite{Basith1}. The first step was the synthesis of MgO nanoparticles, for which 1 mmol Mg(NO\textsubscript{3})\textsubscript{3}.6H\textsubscript{2}O were dissolved in 50 ml of 5M NaOH solution and magnetically stirred for about 4 hours to form a homogeneous mixture. The solution was then transferred into a Teflon-lined autoclave and heated at 180 $^{\circ}$C for 12 hours. The suspension was allowed to cool down to room temperature naturally. Afterwards,  the final products were washed several times with deionized water and ethanol followed by centrifugation, and dried in the oven for 10 hours at 100 $^{\circ}$C. In the next step for incorporating Fe on MgO nanoparticles, 1 mmol Mg(NO\textsubscript{3})\textsubscript{3}.6H\textsubscript{2}O and 0.03 mmol Fe(NO\textsubscript{3})\textsubscript{3}.9H\textsubscript{2}O were dissolved in 50 ml of 5M NaOH solution and the rest of the procedure was similar to that described earlier. After the hydrothermal process, both MgO and Fe/MgO were treated at various temperatures at Ar atmosphere for further characterization.

\begin{figure*}
 \centering
 \includegraphics[width= 1\textwidth]{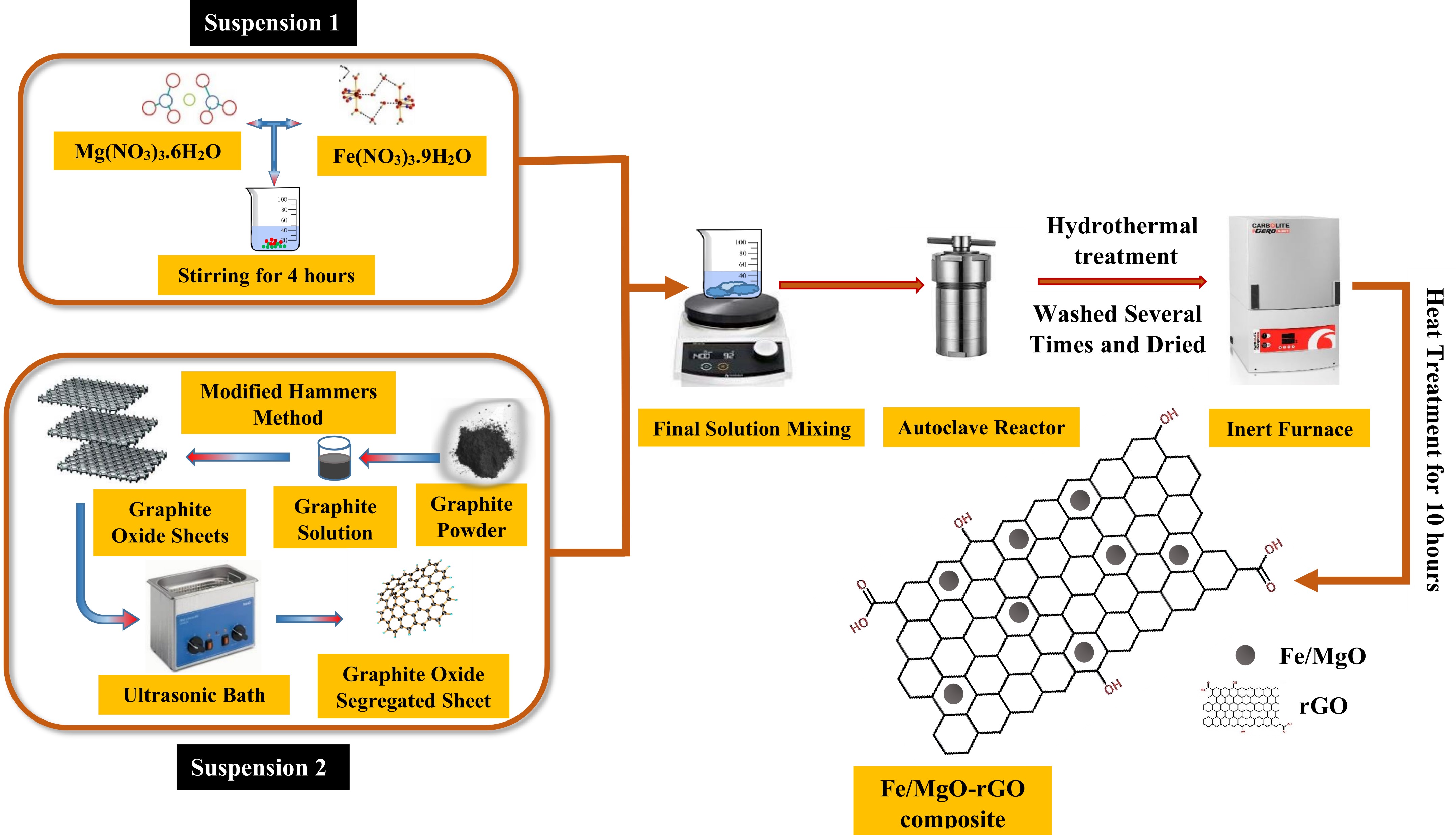}
 \caption{A schematic diagram of the synthesis scheme for rGO supported Fe/MgO nanocomposites}
\end{figure*}

\subsubsection{Synthesis of Fe/MgO-rGO nanocomposites}

Initially, 1 mmol Mg(NO\textsubscript{3})\textsubscript{3}.6H\textsubscript{2}O and 0.03 mmol Fe(NO\textsubscript{3})\textsubscript{3}.9H\textsubscript{2}O  were dissolved in 50 ml of 5M NaOH solution and stirred for 4 hours to get suspension I. In the next step, graphite oxide, prepared by modified Hummer’s method \cite{Paulchamy}, was sonicated for 2 hours in DI water to obtain graphene oxide (GO). Then 10 mg of GO was dispersed in 10 ml DI water to make suspension II . The suspension I and II were combined and stirred for an additional 1 hour. The mixture was then transferred into a Teflon-lined autoclave and heated at 180 $^{\circ}$C for 12 hours. The hydrothermally synthesized products were washed and dried, and we have obtained the final composites after heat treatment of the dried products at 400 and 500 $^{\circ}$C at  Ar atmosphere. A schematic representation of the synthesis process has been depicted in Fig. 1.

\subsection{Characterization techniques}

The Crystalline structure and phase purity of the samples were characterized by the powder X-ray diﬀraction (XRD) spectra using a diﬀractometer (PANalytical Empyrean, UK; Cu-$K_{\alpha}$,$\lambda$ = 1.5418 \AA ). Additionally, the Rietveld refinement approach was conducted for detailed analysis of the experimentally acquired XRD data using FullProf Suite Software \cite{Basith1}. The surface morphology were investigated from field emission scanning electron microscopy (FESEM) imaging along with the energy dispersive X-ray (EDX) analysis using a scanning electron microscope (XL30SFEG; Philips, Netherlands) and transmission
electron microscopy (TEM) (Talos F200X; Thermo fisher scientific, USA)
at an operating voltage of 200 kV. With an aim to observe the optical properties, the absorption spectra of the as-synthesized samples were investigated in the range of 200–600 nm using a UV-visible spectrophotometer (UV-2600; Shimadzu, Japan). Steady-state photoluminescence (PL) analysis was carried out using a Spectro Fluorophotometer (RF-6000; Shimadzu, Japan) to study the rate of photoinduced electron-hole recombination during the photocatalytic process \cite{Tama}. 

\subsection{Photocatalytic degradation of dye}

The photocatalytic performance of the synthesized nanocomposites was assessed by monitoring the degradation of rhodamine B (RhB) dye in DI (deionized) water solution without the presence of any sacrificial agents. A 500W Xe lamp (Hamamatsu Photonics, Japan)  having irradiance value of 100 mW cm\textsuperscript{-2} was used as a solar simulator \cite{Basith2}. For each trial, 80 mg of photocatalyst powder was dissolved in 15 mg/L of RhB solution. Prior to irradiation, the solution was magnetically stirred in the dark for 60 minutes to attain an adsorption equilibrium of RhB on the surface of the photocatalysts. After the lamp was turned on, 4 ml of the  solutions were extracted at 1 hour intervals and centrifuged to filter out the catalyst particles. The absorbance of the solution at a specific wavelength was measured using a UV-vis spectrophotometer to investigate the changes in RhB concentration caused by solar irradiation. The photodegradation process was repeated four times under identical experimental procedures to check the stability and reusability of the photocatalysts.

\subsection{Photocatalytic hydrogen evolution}

Photocatalytic hydrogen evolution experiment was conducted in a slurry type reactor, where 20 mg of photocatalyst were mixed with 30 ml DI water and kept under magnetic stirring before being illuminated by a 500W Hg-Xe lamp. Meanwhile, the system was purged with argon gas for 30 minutes to ensure an inert atmosphere for the splitting process. The evolved hydrogen (H\textsubscript{2}) was extracted at one-hour intervals under illumination, then analyzed and quantified using the GC (Shimadzu, Japan) device equipped with the thermal conductance detector (TCD) and gas analyzer. The GC programming was reverse polarized in order to attain hydrogen peaks in the upward direction to allow a comparative analysis with the peak intensities of the various gases produced \cite{Basith2}.

\section{Results and discussions}

\subsection{Structural characterization}
\subsubsection{Crystal structure}
The XRD patterns of GO, Mg(OH)\textsubscript{2}  and  Fe/Mg(OH)\textsubscript{2}-GO are shown in ESI (electronic supporting information) Fig. S1. The preparation of GO was confirmed from the characteristic peak (002) at Bragg position around 12 $\theta$ \cite{Jalil}, as shown in ESI Fig. S1(a). During the  preparation of MgO following the hydrothermal method, hexagonal Mg(OH)\textsubscript{2} were formed at 180 $^{\circ}$C temperature by redox reactions \cite{Devaraja}. Fig. S1(b) and (c) inserted in ESI demonstrate that the incorporation of Fe and GO into MgO during hydrothermal treatment resulted in the formation of Fe/Mg(OH)\textsubscript{2}-GO at  the same temperature. In the XRD pattern of Fe/Mg(OH)\textsubscript{2}-GO, we observe a peak for GO which indicates that the GO was not reduced. This can be attributed to the absence of any chemically reducing agent during hydrothermal preparation.\\
\tab The Rietveld reﬁnement method was used for detailed structural analysis of the samples after heat treatment at Ar atmosphere. Fig. 2(a), (b), (c), and (d) display the Rietveld reﬁned powder XRD patterns of synthesized materials after heat treatment at Ar atmosphere. MgO and Fe/MgO nanoparticles heated at 500 $^{\circ}$C are represented by MgO(500) and Fe/MgO(500), and Fe/MgO-rGO nanocomposites heated at 400, and 500 $^{\circ}$C are represented by Fe/MgO-rGO(400) and Fe/MgO-rGO(500), respectively. It is confirmed that following the heat treatment at 500 $^{\circ}$C at Ar atmosphere, the synthesized hexagonal structured Mg(OH)\textsubscript{2} was converted into cubic MgO which is consistent with a previous investigation \cite{Devaraja}.
\begin{figure}[h!]
 \centering
 \includegraphics[width= 8.5 cm]{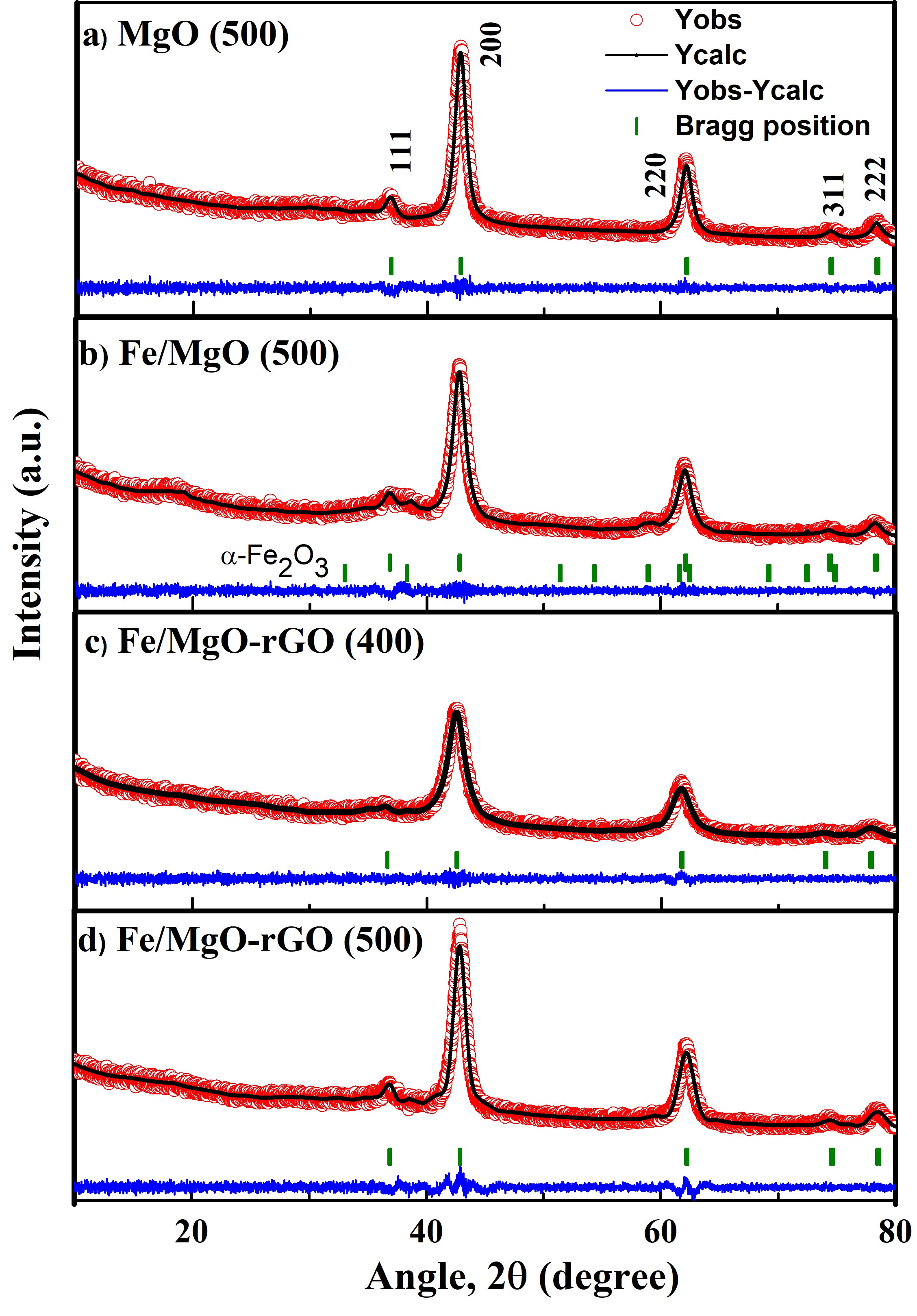}
 \caption{XRD patterns of (a) MgO(500), (b) Fe/MgO(500) nanoparticles, and (c) Fe/MgO-rGO(400), (d) Fe/MgO-rGO(500) nanocomposites after heat treatment at Ar atmosphere}
\end{figure}

Moreover, it can be observed that all samples display characteristics diffraction peaks indexed at (111), (200), (220), (311) and (222) planes which are in good accordance with cubic MgO with \textit{Fm-3m} space group (JCPDS card no. 1-75-447 \cite{Petnikota}). Fig. 2(a) and (b) show that due to the incorporation of Fe in MgO(500) nanoparticles, the peak positions remained the same, which suggested that Fe did not alter MgO crystal structure and was suitably deposited on the surfaces of MgO \cite{Borhade}. Though the presence of an additional iron oxide phase $\alpha$-Fe\textsubscript{2}O\textsubscript{3}  was observed in Fe/MgO(500) samples (Fig. 2(b))  due to iron substitution, which was eliminated when GO was added into the sample and heated at inert atmosphere as shown in Fig. 2(c) \cite{Pin}.\\

\tab Comparing the XRD patterns of ESI Fig. S1(c) and Fig. 2(c), the formation of Fe/MgO-rGO is assured  at 400 $^{\circ}$C  temperature with an absence of GO peak indicating the total reduction of GO. This is consistent with the reduction mechanism of GO by thermal deoxygenation, in which an increase in temperature facilitates the removal of oxygen-containing functional groups from GO \cite{Ganguly,Jin}. This also explains the disappearance of $\alpha$-Fe\textsubscript{2}O\textsubscript{3} phase due to GO substitution. With the increased temperature at Ar atmosphere, slightly decreased peak broadening as well as an increase in peak intensities was observed (Fig. 2(c) and (d)). This decreasing trend of peak broadening indicates the increase in particle size, suggesting agglomeration at a higher temperature \cite{Kayani, Sinornate}.\\  
\tab Table 1 presents diﬀerent structural parameters along with constituent phases for as-synthesized samples. Notably, the calculated lattice parameters are well consistent with the values reported in the literature \cite{Devaraja1}. The lattice parameters and volume corresponding to the MgO phase remain almost unaltered throughout the synthesis process, which confirmed that the crystal structure was not distorted. The refined and observed XRD patterns for the cubic MgO phase are in reasonable agreement according to the fitting parameters (R\textsubscript{wp}, R\textsubscript{p}, R\textsubscript{Exp} and $\chi^{2}$) and atomic coordinates (ESI Table S1).

\begin{table*}[t!]
\small
\caption[centering]{Structural parameters of synthesized nanomaterials and nanocomposites.}
\begin{tabular*}{\textwidth}{@{\extracolsep{\fill}}llllllll}

\hline
Sample & Constituent & Crystallographic phase & Space group & a = b (\AA) & c (\AA) & V (\AA\textsuperscript3)                          \\ \hline
MgO (500) &MgO & Cubic & \it{F m -3 m} &   4.220           & 4.220 & 75.176(0.005) \\ & & & & \\ 
Fe/MgO (500) & MgO & Cubic & \it{F m -3 m}  &  4.227 & 4.227 & 75.576(0.005)\\ 
&         $\alpha$-Fe\textsubscript{2}O\textsubscript{3} & Trigonal & \it{R -3 c}
& 5.430 & 5.429 &138.649(0.001) \\ & & & & \\
Fe/MgO-rGO (400) & MgO & Cubic & \it{F m -3 m} &  4.250        & 4.250 & 76.794(0.008) \\ & & & & \\ 
 Fe/MgO-rGO (500) & MgO  & Cubic & \it{F m -3 m} & 4.221 & 4.221 & 75.243(0.005) 
\\ 
  \\\hline
\end{tabular*}
\end{table*}

\subsubsection{Surface morphology}
 The FESEM image in Fig. 3(a) demonstrates randomly aggregated, thin, crumpled sheets like structure of GO \cite{Jalil}. In  Fe/Mg(OH)\textsubscript{2}-GO samples, without heat treatment (ESI Fig. S2(a)), numerous Fe/Mg(OH)\textsubscript{2} hexagons with different particle sizes were predominately observed to be grown and randomly distributed on the GO surface. 
\begin{figure*}
 \centering
 \includegraphics[width= 14 cm]{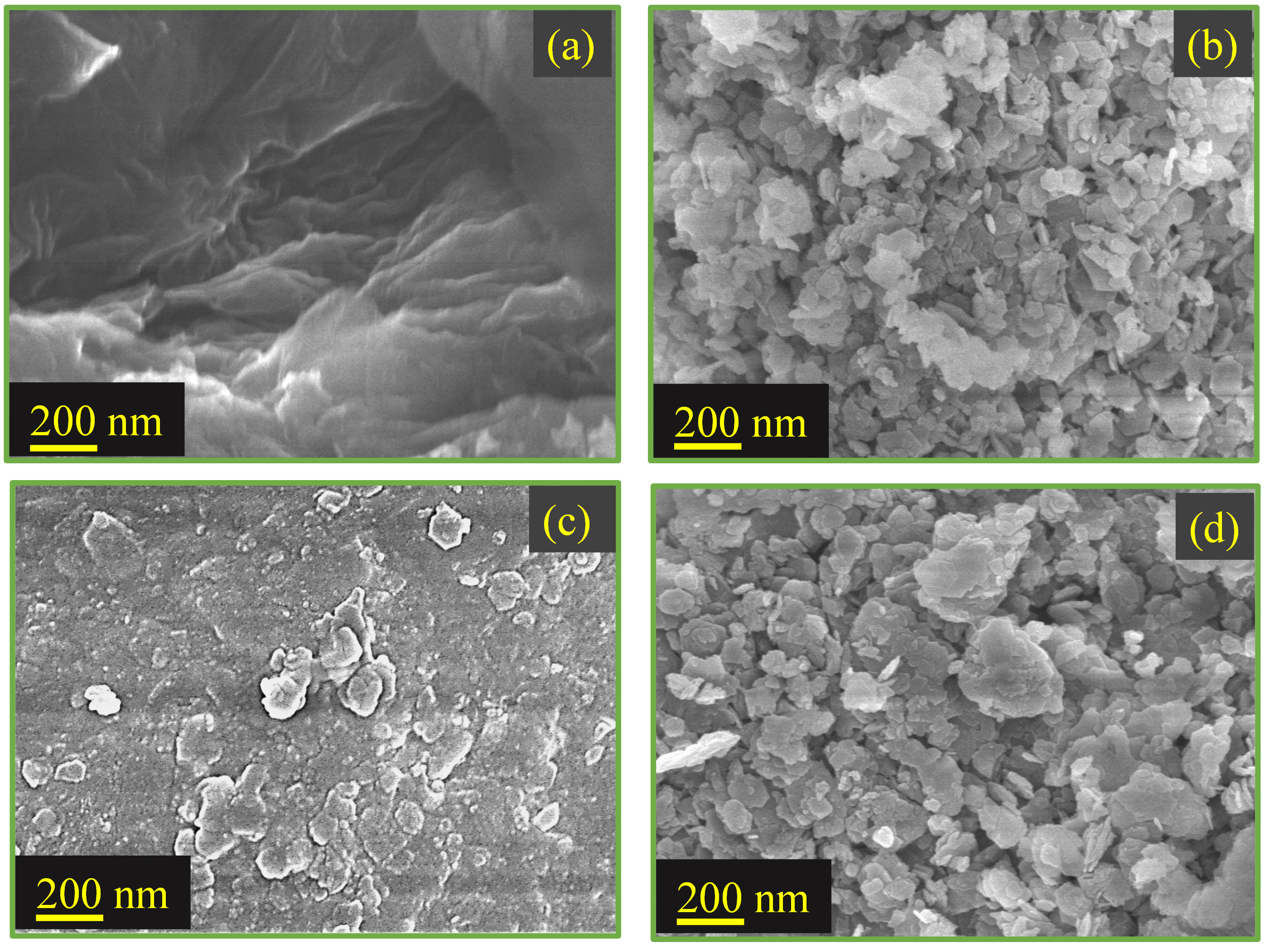}
  \caption{FESEM images of (a) GO, (b) Fe/MgO(500), (c) Fe/MgO-rGO(400),and (d) Fe/MgO-rGO(500) } 
\end{figure*}
Fig. 3(b), (c), and (d) represent the FESEM images of Fe/MgO sample heated at 500 $^{\circ}$C and Fe/MgO-rGO samples heated at 400 and 500 $^{\circ}$C at Ar atmosphere, respectively. It can be seen that for Fe/MgO-rGO(400) sample, the surface are not homogeneous, and the particles growth is not up to the mark, which can be ascribed to the fact that hexagonal Fe/Mg(OH)\textsubscript{2} did not completely transformed into cubic Fe/MgO at lower heating temperature (400 $^{\circ}$C). However, due to graphene incorporation, a large number of  Fe/MgO nanocrystallites nucleate and grow into fine seed particles at the early stage of hydrothermal treatment; with time, this is followed by the growth of Fe/MgO nuclei \cite{Turabik, Li}. Surprisingly, when composites are formed with rGO, such consumption of finer seed particles that lead to the growth of nuclei is interrupted due to steric blockage of seed particle transportation caused by the interpenetrating graphene networks \cite{WLi}. However, for  Fe/MgO-rGO nanocomposite at 500 $^{\circ}$C temperature, the particle size distribution seems to be more homogeneous than the samples treated at 400 $^{\circ}$C. For a further increase in temperature  to 600 $^{\circ}$C, resulted in a significant amount of agglomeration \cite{KumarA}, as can be observed in ESI Fig. S2(b). This result implies that 500 $^{\circ}$C is the optimized Ar atmosphere required to form Fe/MgO-rGO nanocomposites with a higher degree of uniformity in shape and size. To get further insight into the morphology of 
the rGO supported Fe/MgO nanocomposite at 500 $^{\circ}$C, TEM imaging was performed at various magnifications (Fig. 4). The TEM image, Fig. 4(a) clearly demonstrated the distribution of Fe/MgO nanoparticles on rGO sheets. The higher magnification images, Fig. 4(b) and (c) confirmed the presence of graphene sheets. In Fig. 4(d) we have observed bright concentric rings in the selected area electron diffraction (SAED) pattern, which implied the polycrystalline nature of the synthesized nanocomposite.

\begin{figure*}
 \centering
 \includegraphics[width= 0.75\textwidth]{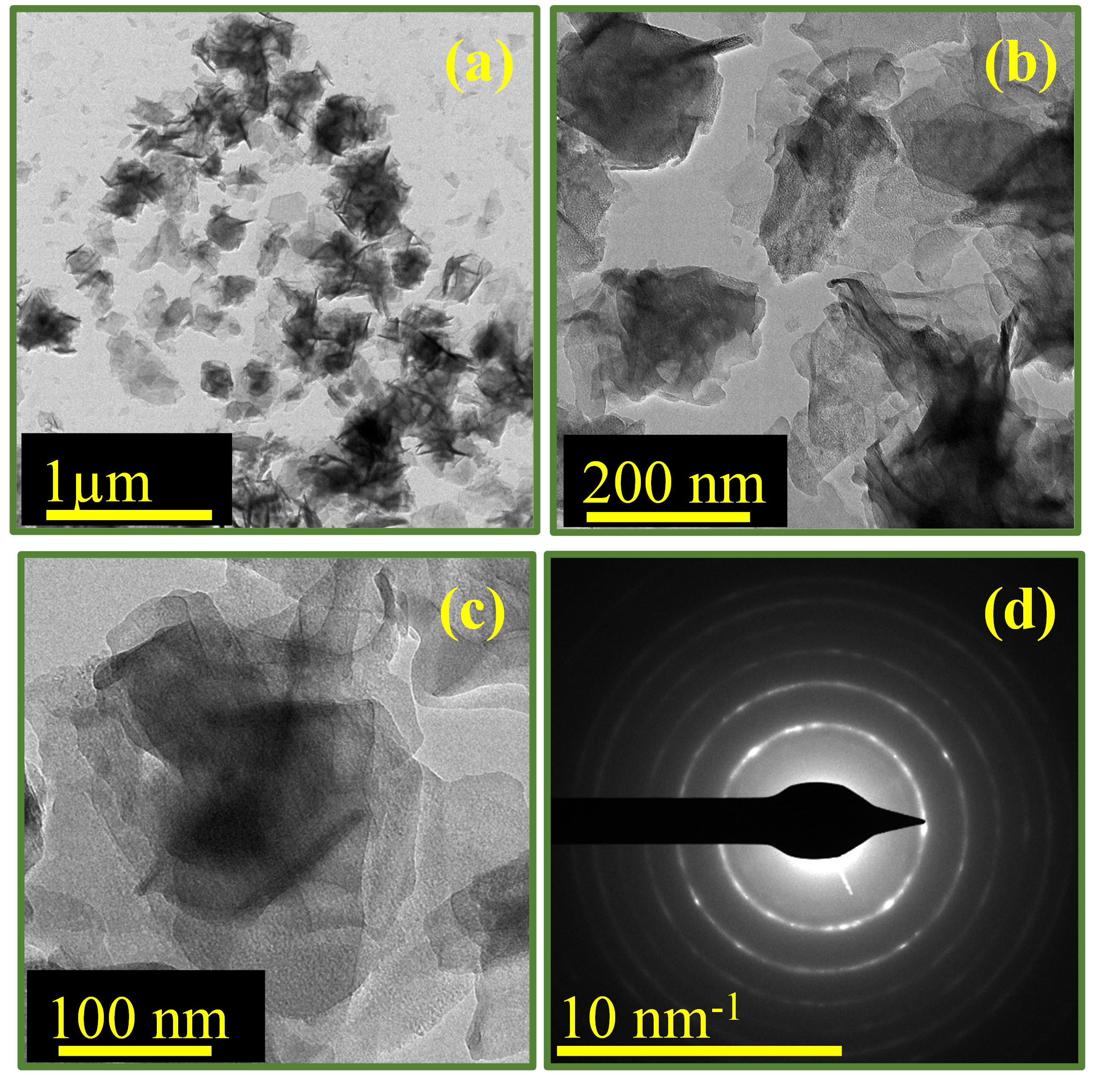}
  \caption{TEM images (a-c), and SAED pattern (d) of Fe/MgO-rGO(500) nanocomposite } 
\end{figure*}

\subsubsection{Elemental composition}
The compositions of the synthesized samples were identified utilizing EDX analysis. The experimental details in ESI Fig. S3 show that the desired elements C, Mg, O and Fe are present in the synthesized samples, and no additional element was found. The EDX spectra were taken at three distinct positions, and the average atomic ratios of Mg, Fe, O and C were observed to be very close to nominal composition within the instrumental accuracy.

\subsection{Optical characterization}
\subsubsection{Optical band gap}

The optical properties of all the synthesized samples were studied using UV-visible diffuse reflectance spectroscopy (DRS) in the wavelength range of 200–600 nm. Fig. 5(a) demonstrates the room temperature UV-visible absorption spectra of pure MgO, Fe/MgO nanoparticles and Fe/MgO-rGO nanocomposites after heat treatment at different reaction temperatures. All the samples exhibited continuous absorption across the UV region, at the same time, the addition of rGO showed an increase in the optical absorption compared to pure MgO and Fe/MgO samples, indicating higher light absorption ability of the synthesized nanocomposites. Moreover, MgO nanoparticles exhibit a prominent absorption peak at 286 nm, as shown in Fig. 5(a). In contrast all Fe-incorporated samples show a blue shift in the absorption peak, which can be attributed to the formation of donor energy levels in the actual band gap of MgO. \\
\tab The optical band gap $E_{g}$ of the samples were calculated using Tauc’s relation \cite{Basith2} 
\begin{equation}
    F(R) \times h\nu=A(h\nu-E_{g})^{n}
\end{equation}
where $h\nu$, A and $E_{g}$  denote the energy of the photon, proportionality constant and optical band gap, respectively and F(R) is a parameter calculated from DRS by Kubelka–Munk function \cite{Basith1}.
From Fig. 5(b), the band gap energy was determined by extrapolating the straight-line portion to the abscissa at zero absorption co-efficient. The band gap of MgO  is found to be 3.9 eV which has been reduced to 3.3 eV when Fe was incorporated into MgO. Incorporation of rGO suppressed the crystal growth, resulting in smaller particle size and the band gap energy of Fe/MgO-rGO(400) nanocomposite was reduced to 2.78 eV \cite{Jalil}. As the temperature rises from 400 to 500 $^{\circ}$C, there is an increase in band gap from 2.78 to 3.0 eV due to quantum confinement effect \cite{Nanakkal}. Another reason is that extremely high temperature promotes the generation of strong bonds between the crystallites, leading to the agglomeration of particles \cite{Zhang}. Interatomic bonds are tightened and high energy is required to break a bond, and move an electron to the conduction band. 

\begin{figure*}
 \centering
 \includegraphics[width= 16 cm]{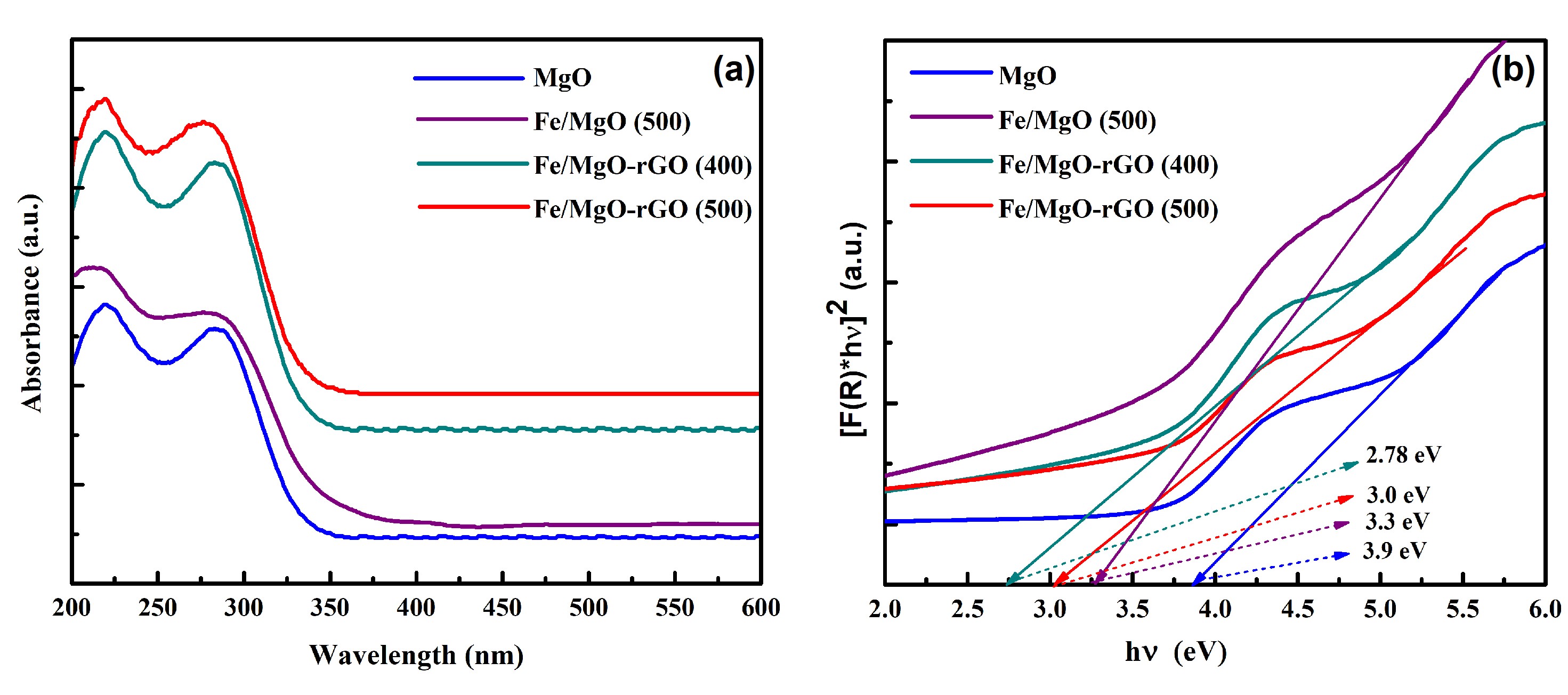}
 \caption{ (a) UV-visible absorbance and (b) Tauc plots for band gap estimation for MgO, Fe/MgO nanoparticles heated at 500 $^{\circ}$C and Fe/MgO-rGO  nanocomposites heated at 400 and 500 $^{\circ}$C at Ar atmosphere}
\end{figure*}

\subsubsection{Photoluminescence (PL) spectra at steady-state}
A photocatalyst must possess the ability to inhibit the electron-hole pair recombination efficiently. To study this property, we have conducted the PL analysis of our samples for an excitation wavelength of 200 nm as represented in ESI Fig. S4. The optical band gaps from the position of the PL peaks are observed to be 3.40, 2.69 and 3.03 eV for Fe/MgO(500), Fe/MgO-rGO(400) and Fe/MgO-rGO(500) samples, respectively. These values are in accordance with the band gaps obtained from UV-visible spectroscopy (Fig. 5(b)). Notably, the peak intensity of both rGO supported samples are lower than Fe/MgO samples, which suggests that the separation of photogenerated electron-hole pairs can be facilitated through rGO incorporation. This result is also consistent with our previous band gap estimation by DRS, where the band gaps of all rGO assisted nanocomposites are smaller than those of Fe/MgO.\\
\tab Moreover, for Fe/MgO-rGO(500), the peak intensity is lowest as compared to other samples, which imply the successful separation and transfer of photogenerated charge carriers. As a result, the photocatalytic activity of this nanocomposite under solar irradiation increases by inhibiting electron-hole pair recombination \cite{Ghosh}. The outcome of our experiment is also in good agreement with previous reports, where PL emission is observed to be  quenched due to rGO substitution \cite{He,Hareesh}

\subsection{Photocatalytic degradation of rhodamine B (RhB) dye}

\begin{figure*}[t!]
\centering
\includegraphics[width=14 cm, height=18 cm]{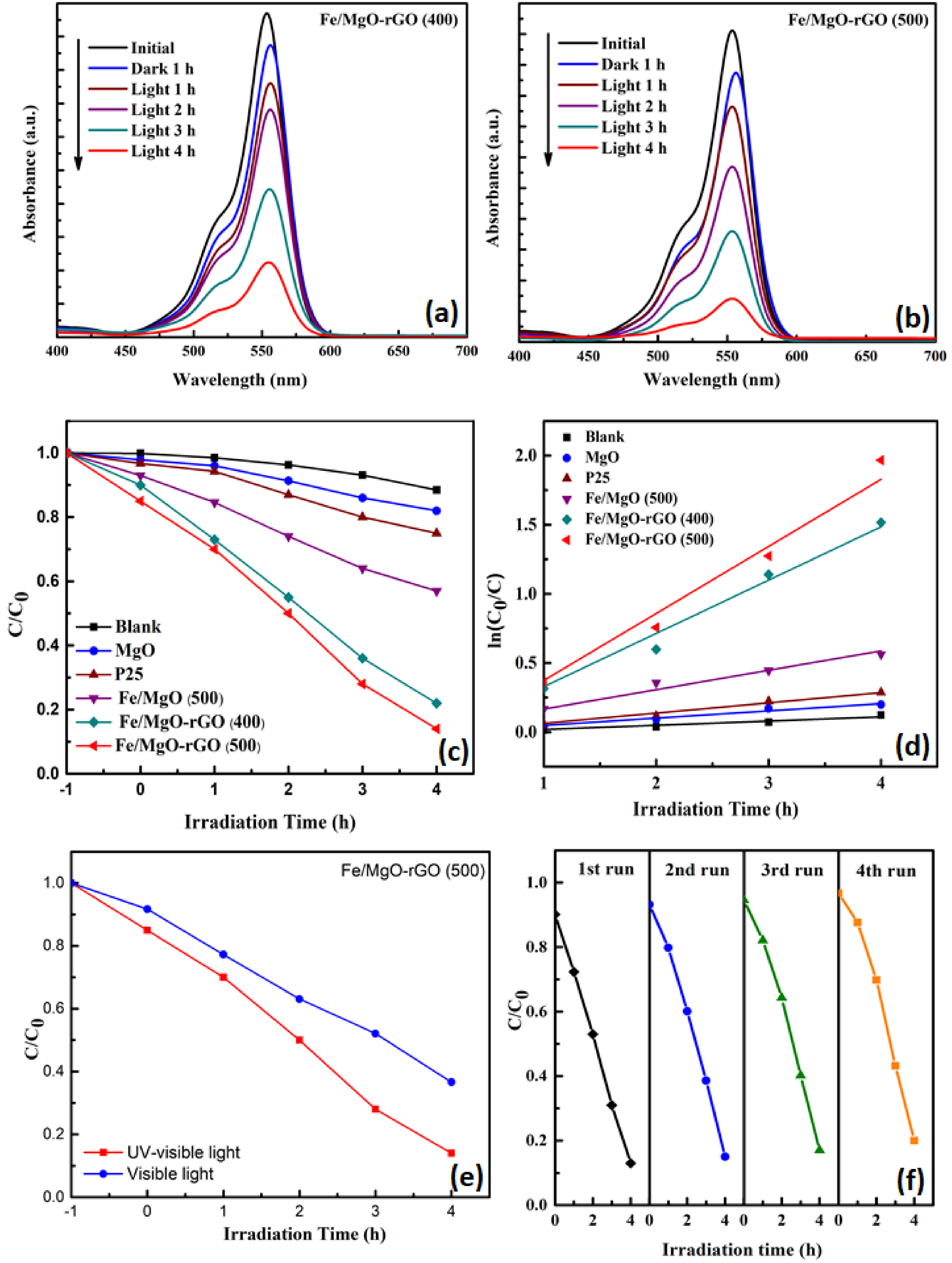}
\caption[centering]{Spectral changes during the deterioration of RhB over (a) Fe/MgO-rGO(400), and (b) Fe/MgO-rGO(500) nanocomposites under solar irradiation;(c) RhB concentration change as a function of irradiation time, (d) First-order kinetic plots for the photodegradation of RhB, (e) Change in RhB concentration under different wavelength conditions, and (f) Recyclability test of Fe/MgO-rGO(500) for four successive runs}
\end{figure*}
To evaluate the photocatalytic performance of the as-synthesized materials, at first, a blank test was conducted without any catalyst. RhB did not show significant degradation for four hours indicating the self-degradation ability of RhB is extremely low. After that, the catalyst samples were blended with RhB solution and kept in the dark under constant magnetic stirring in order to achieve adsorption equilibrium. In Fig. 6(a) and (b), we have  presented the time-dependent UV-visible absorbance spectra of RhB at one-hour intervals in the presence of Fe/MgO-rGO(400) and Fe/MgO-rGO(500) photocatalysts. The decrease in absorbance peak  as irradiation time extends, indicates the decomposition of RhB under solar irradiation. It was observed that incorporation of rGO, and heat treatment has resulted in a significant improvement in degradation efficiency of the nanocomposites and the maximum degradation of RhB was observed at 500 $^{\circ}$C at Ar atmosphere. This result may be attributed to the relationship between the stability of the crystal structure and the effectiveness of reduced graphene oxide for solar light absorption \cite{Zhang}.\\
\tab The degradation efficiency as well as degradation rate were calculated by using the following equation
\begin{equation}
    Photocatalytic\;\;degradation (\%)= \frac{C_{0}-C}{C_{0}}\times 100
\end{equation}
Fig. 6(c) demonstrates the degradation characteristics of RhB as a function of solar irradiation time in the presence of as-synthesized MgO, Fe/MgO(500), Fe/MgO-rGO(400) and Fe/MgO-rGO(500) photocatalysts as well as commercially available Degussa TiO\textsubscript{2} (P25) nanoparticles for comparison. 
C\textsubscript{0} and C represent the primary and retained concentrations of RhB in the solution, respectively. The degradation efficiency of pure MgO, Fe/MgO(500), Fe/MgO-rGO(400) and Fe/MgO-rGO (500) was $\sim$25\% , $\sim$42\%, $\sim$80\% and $\sim$86\%, respectively. The photodegradation curve of commercial titania (P25) was also plotted in Fig. 6(c) for comparison. Interestingly, the degradation percentage of all the samples except MgO is higher compared to P25. The large band gap and quick charge carrier recombination might be responsible for this poor degradation efficiency of MgO. Due to the presence of Fe on MgO, the band gap was reduced, the active surface of MgO was increased, and the RhB dye degradation efficiency was enhanced accordingly \cite{Ot, Man}. Moreover, for comparison we have also evaluated the photodegradation ability of  Fe\textsubscript{2}O\textsubscript{3}, Fe\textsubscript{2}O\textsubscript{3}-MgO, and Fe\textsubscript{2}O\textsubscript{3}-rGO samples for comparison.  It was observed that under four hours of solar irradiation, the degradation efficiency was  $\sim$38\%, $\sim$43\%, and $\sim$52\% for the respective samples, which was significantly lower than that of Fe/MgO-rGO nanocomposites. The synthesis procedure of the above-mentioned samples and the experimental findings of the photodegradation experiments have been attached in the supplementary (ESI Fig. S5). From these observations, it can be elucidated that the inclusion
 of rGO in Fe/MgO-rGO nanocomposite significantly improved their photocatalytic performance. The highest photocatalytic degradation efficiency of $\sim$86\% was observed for Fe/MgO-rGO nanocomposite, which was heated at 500 $^{\circ}$C  at Ar atmosphere. We think that at optimum temperature of 500 $^{\circ}$C maximum number of oxides were removed from GO and resulted in the successful migration of electrons from the conduction band (CB) of Fe/MgO to rGO \cite{Arshad}. This migration inhibited electron-hole pair recombination and resulted in an outstanding photocatalytic performance of Fe/MgO-rGO(500) nanocomposite \cite{Gayathir}. In addition, from the FTIR analysis of Fe/MgO-rGO(500) nanocomposite we can observe peaks at 1450 and 3390 cm\textsuperscript{-1}, indexed to –OH and H–O–H vibration bonds, respectively (ESI Fig. S6). These oxygen-containing groups have an effective role in the photocatalytic activity of catalysts by generating more hydroxyl radicals to facilitate the photocatalytic process \cite{Arshad,Isari}.\\
\tab We have incorporated the Pseudo first-order kinetics model to quantitatively study the kinetics of the photocatalytic degradation using the following equation\cite{Sub}: 
\begin{equation}
    ln (C\textsubscript{0}/C)=kt
\end{equation}
Where C\textsubscript{0} and C represent the same as previously stated, and k (min \textsuperscript{-1}) is the pseudo-first order rate constant of photodegradation. The rate constant was calculated by using a linear fit to the plot of ln(C\textsubscript{0}/C) versus irradiation time. Based on the results from Fig. 6(d), the highest degradation rate was observed for Fe/MgO-rGO(500) with a rate constant of k=7.072×10\textsuperscript{-3} min\textsuperscript{-1}, which is 2.86 times higher than that of Fe/MgO (k=2.479×10\textsuperscript{-3} min\textsuperscript{-1}). Considering the degradation efficiency and rate constant, Fe/MgO-rGO(500) can be regarded as a superior solar light driven photocatalyst to be used for environmental remediation purpose. The mechanism behind this profound performance of Fe/MgO-rGO(500) has been discussed subsequently.\\
\tab Furthermore, to study the role of light with different wavelengths, the photodegradation ability of Fe/MgO-rGO(500) nanocomposite was also evaluated using a UV cut-off filter ($\lambda\geq$ 420 nm) where the efficiency was obtained to be only $\sim$64\% (Fig. 6(e)) which is lower than that obtained under UV-visible irradiation. This result justifies our findings from optical characterizations, where continuous absorption across the UV region was exhibited by Fe/MgO-rGO composites.

\subsubsection{Photostability of Fe/MgO-rGO(500) nanocomposite}
The stability and re-usability of the Fe/MgO-rGO(500) nanocomposite has been checked, employing a recyclability test to validate their applicability in practical fields \cite{Basith1}. It has been observed from Fig. 6(f)  that after four consecutive photo-degradation cycles, the change in dye degradation efficiency of the catalyst was trivial, which could be related to the loss in material during the recovery process. The outcome of this study anticipates the prospect of this material for repeated application in waste-water management and other photocatalytic purposes.

\subsection{Photocatalytic hydrogen generation}
 We have evaluated the efficiency of the synthesized photocatalysts as a worthy candidate for hydrogen generation via water splitting under solar irradiation \cite{Basith2}. Fig. 7 demonstrates the amount of H\textsubscript{2}  gas produced by our synthesized catalysts as a function of solar irradiation time. For comparison, the hydrogen generation under identical conditions by titania P25 nanoparticles was also included. The hydrogen production of commercially available P25 nanoparticles was observed to be $\sim$445 $\mu$mol g\textsuperscript{-1}. As shown in Fig. 7, Fe/MgO photocatalyst's hydrogen evolution is slightly higher than that of P25 nanoparticles. Notably, the rate of hydrogen production increased significantly when rGO was introduced into Fe/MgO, and it was observed that the total amount of H\textsubscript{2}  evolved under 4 hours of solar irradiation were $\sim$1517 and $\sim$1741 $\mu$mol g\textsuperscript{-1} for the Fe/MgO-rGO(400) and Fe/MgO-rGO(500) nanocomposites, respectively.
\begin{figure}
 \centering
 \includegraphics[width= 0.47\textwidth]{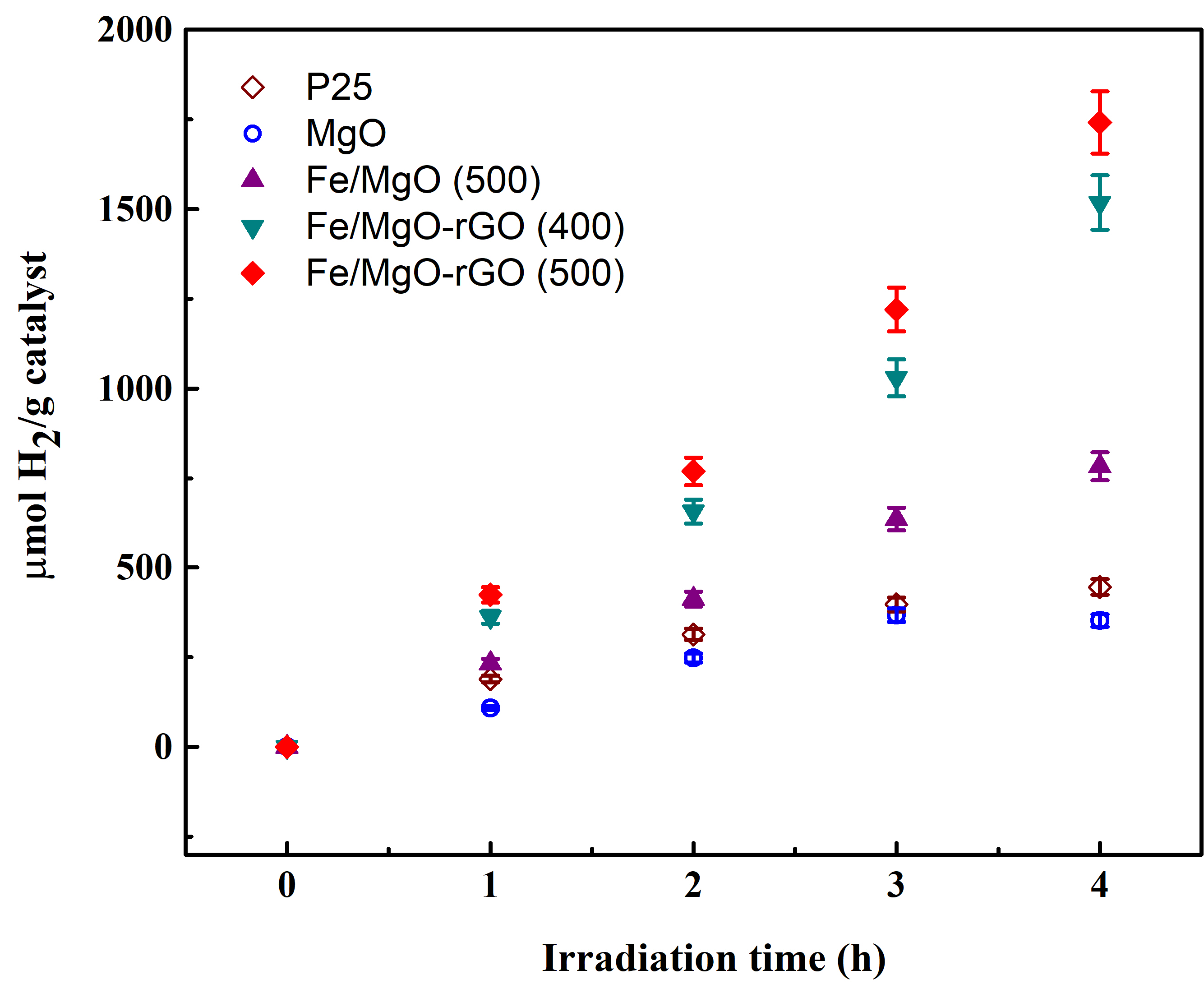}
 \caption{Amount of  H\textsubscript{2} evolved via water
splitting with respect to irradiation time 
}
\end{figure}

It is well established that Photocatalytic H\textsubscript{2}  generation mainly includes three basic steps: (1) absorption of photons by the photocatalyst to form electron-hole pairs; (2) separation and migration of the photo-induced charges to different photocatalyst sites; and (3) reduction of water to evolve H\textsubscript{2} on the surface of the photocatalyst by the photo-generated electrons. In our case, Fe/MgO-rGO(500) nanocomposite  absorbed a significant amount of photons to produce electron-hole pairs due to its narrow band gap and graphene, with its high work function (4.42 eV) and superior conductivity, played the crucial role by accepting photo-generated electrons and rapidly migrating these electrons through its 2D plane to reactive sites, which resulted in a significant amount of H\textsubscript{2}  evolution \cite{Liu,Wang}. For  further insight, a comparative analysis between our experimental findings and some previous related investigations has been presented in Table 2. 
Noticeably, our Fe/MgO-rGO(500) nanocomposite exhibited a comparatively increased   H\textsubscript{2}   evolution rate (HER) than other rGO supported and MgO based nanocomposites. Additionally, in most cases, various sacrificial reagents were used as electron donors to enhance H\textsubscript{2} generation rate. However, in this present investigation, we have obtained a notable result on solar hydrogen generation by water splitting without any additional reagent.

\begin{table*}[]

\caption{An overview  of  the  photocatalytic  H\textsubscript{2}   production  ability of  photocatalysts  reported in recent literature.}
\resizebox{\textwidth}{!}
{\begin{tabular}{lllll}
\hline
Photocatalyst                  & Light source                          & Reactant solution                        & H$_{2}$ \;\;yield       & Reference (year)    \\
                               &                                       &                                             & ($\mu$  mol g$^{-1}$ h$^{-1})$ &              \\ \hline
SrTiO\textsubscript{3}-rGO & 300 W Xe lamp  & H\textsubscript{2}PtCl\textsubscript{6}& 363              & \cite{He} (2016)           \\
ZnO/rGO & 300 W Xe lamp                         & Triethylamine, alcohol& 616          & \cite{Kang} (2016)      \\
TiO\textsubscript{2}-rGO & 300 W Xe lamp; $\lambda\textgreater{}$300\;\;nm & Methanol       & 720          & \cite{El} (2017)
\\
TiO\textsubscript{2}/Pt/rGO& 4 Philips PL-S (9 W) lamp; 315$\textless\lambda\textless{}$400\;\;nm& Methanol & 259            & \cite{Rivero} (2018)           \\
MgO/MgCr\textsubscript{2}O\textsubscript{4} & Hg lamp; $\lambda\geq$ 400\;\; nm                   & Methanol            & 420            & \cite{Nayak} (2018)         \\
Bi\textsubscript{25}FeO\textsubscript{40}-rGO & 500 W Xe lamp; $\lambda \textless{}$400\;\;nm & Deionized water               & 390          & \cite{Basith2}  (2018)        \\
CdS/rGO                   & 300 W Xe lamp; $\lambda\textgreater{}$420\;\;nm & Na\textsubscript{2}S, Na\textsubscript{2}SO\textsubscript{3}& 500           & \cite{Hareesh} (2019)           \\
TiO\textsubscript{2} /rGO/LaFeO\textsubscript{3} & 300 W Xe lamp                   & Methanol & 893       & \cite{Lv}  (2019)         \\
PANI-TiO\textsubscript{2} /rGO & 300 W Xe lamp                      & Triethanolamine & 806           & \cite{Ma} (2020)           \\
TiO\textsubscript{2} /In\textsubscript{0.5}WO\textsubscript{3}-rGO                   &  Xe arc lamp   &Glycerol  &304        & \cite{Shaheer} (2021)           \\
Fe/MgO-rGO& 500 W Xe lamp & Deionized water               & 432          & Present work        \\
\hline
\end{tabular}}
\end{table*}

\subsection{Photocatalytic mechanism}
In order to gain a better perspective of the profound photocatalytic performance of Fe/MgO-rGO nanocomposites, a plausible mechanism for the photodegradation of RhB under solar illumination is proposed with the help of a schematic illustration (Fig. 8).  The band edge positions were calculated using the Mulliken electronegativity method using the following equations \cite{Nethercot}
\begin{equation}
    E_{CB}=\chi -E_{c}-\frac{1}{2}E_{g}
\end{equation}
\begin{equation}
    E_{VB}=E_{CB}+E_{g}
\end{equation}
where $E_{CB}$ and $E_{VB}$ are conduction and valence band edge potentials with respect to Normal Hydrogen Electrode (NHE) ; $\chi$ is the absolute electronegativity, expressed as the geometric mean of the absolute electronegativity of the constituent atoms; $E_{g}$ is the band gap of the sample and $E_{c}$ is the energy of free electrons on the hydrogen scale ($\sim$4.5 eV). It is well established that, along with optical band gap, CBM (conduction band minimum) and VBM (valance band maximum) play crucial roles  in catalyzing the redox reactions in the electrolyte system. Efficient photocatalytic degradation ability of the nanocomposites can be better understood by analyzing the redox reactions in the electrolyte system that govern the degradation of RhB dye. For successful degradation of RhB under solar irradiation two redox reactions are particularly important. First, the photogenerated electron can react with the surface adsorbed oxygen
(redox potential: -0.16 eV vs. NHE) to form $\cdot O_{2}^{-}$ radicals, which will further react with RhB to cause degradation. Second, 
the photogenerated holes can react with the $OH^{-}$ radicals, ionized from the water molecules to produce $\cdot OH$ (redox potential: 2.38 eV vs. NHE). This $\cdot OH$ can further oxidize the RhB molecules. Therefore, a photocatalyst needs to possess 
a CBM\textless{}-0.16 eV  to drive the frst reaction and VBM\textgreater{}2.38 eV to drive the second one efficiently \cite{Jalil}.

CBM and VBM for MgO has been determined to be -1.24 and 2.66 eV, respectively, which imply that MgO can perform the redox reactions of RhB degradation since it has fulfilled the above mentioned criteria. However, lower absorption of photons due to the large value of optical  band gap in MgO, resulted in poor photocatalytic activity. Introduction of iron Fe\textsuperscript{3+} ions into the MgO surface,  effectively leads to band gap narrowing, creating a sub level above the valance band (VB) of MgO. Under solar illumination, electrons in the VB of MgO can firstly transfer to Fe intermediate band through absorbing photons with longer wavelength, and these electrons then transfer from intermediate band to conduction band (CB) leaving  holes in the VB. Moreover, the Fe\textsuperscript{3+} ions can be reduced to Fe\textsuperscript{2+} ions by the photogenerated electrons during the photocatalysis,  which greatly improves the charge separation with increased lifetime of excited electrons and holes \cite{LiJ,Liang}. 
Fe\textsuperscript{2+}  ions are relatively unstable when 
compared to Fe\textsuperscript{3+} ions, which have half-filled 3d\textsuperscript{5} orbital. Therefore, the trapped charges can easily 
release from Fe\textsuperscript{2+} ions and then migrate to the surface to initiate the photocatalytic reaction.
In addition, Fe\textsuperscript{2+} ions can be oxidized to Fe\textsuperscript{3+} ions by transferring electrons to absorbed molecular oxygen in the reaction solution, resulting in the formation of reactive superoxide radicals ($\cdot O_{2}^{-}$) which can 
further degrade RhB \cite{Liang}.\\
\tab The Fe/MgO nanoparticles are supposed to be well attached to the rGO sheets as can be seen in the TEM images. Due to its excellent electron conductivity, graphene can greatly promote interfacial charge transfer and significantly improve the photoactivity of Fe/MgO-graphene nanocomposites. Furthermore, due to the $\pi$–$\pi$ conjugated interactions between the RhB molecules and the graphene rings, graphene sheets have an exceptional ability to adsorb dye molecules in the aqueous medium \cite{Ma}. Besides, density-functional theory (DFT) calculations revealed that the work function (WF) of graphene is 4.42 eV, indicating the redox potential of rGO/rGO$\cdot^{-}$ is (-0.08 eV versus NHE) more positive than the CB potential of Fe/MgO (-1.24 eV versus NHE) \cite{LiQ,Chen}. This will result in a thermodynamically convenient transfer of photogenerated electrons from the CB of  Fe/MgO  to the rGO surface, giving the electrons more opportunities to interact with reactants and participate in photocatalytic reactions. The photo generated electron in rGO could react with adsorbed  O\textsubscript{2} to form $\cdot O_{2}^{-}$ radicals. The majority of $\cdot O_{2}^{-}$  radicals will be involved in the degradation of RhB molecules, while others may react with $H_{2}O$ to produce $\cdot OH$ radicals. Consequently, the produced  $\cdot O_{2}^{-}$ and $\cdot OH$ radicals, as well as photogenerated electrons and holes will react with dye molecules and degrade it into non-toxic byproducts.\\
\tab The tentative mechanism can be summarized as follows
\begin{equation}
Fe/MgO + h\nu \rightarrow Fe(e^{-})/MgO(h^{+})
\end{equation}
\begin{equation}
Fe^{3+}+ e^{-} \rightarrow Fe^{2+}     
\end{equation}
\begin{equation}
Fe^{2+}+ O_{2} \rightarrow Fe^{3+}+\cdot O_{2}^{-}
\end{equation}
\begin{equation}
MgO(h^{+}) + H_{2}O \rightarrow H^{+} +\cdot OH
\end{equation}
\begin{equation}
e^{-} +rGO \rightarrow rGO(e^{-})
\end{equation}
\begin{equation}
rGO(e^{-}) + O_{2}  \rightarrow \cdot O_{2}^{-}
\end{equation}
\begin{equation}
RhB + (\cdot O_{2}^{-}+\cdot OH + e^{-}+ h^{+} ) \rightarrow Degraded\;\;products
\end{equation}

\begin{figure*}
\centering
\includegraphics[width= 16 cm, height=10 cm]{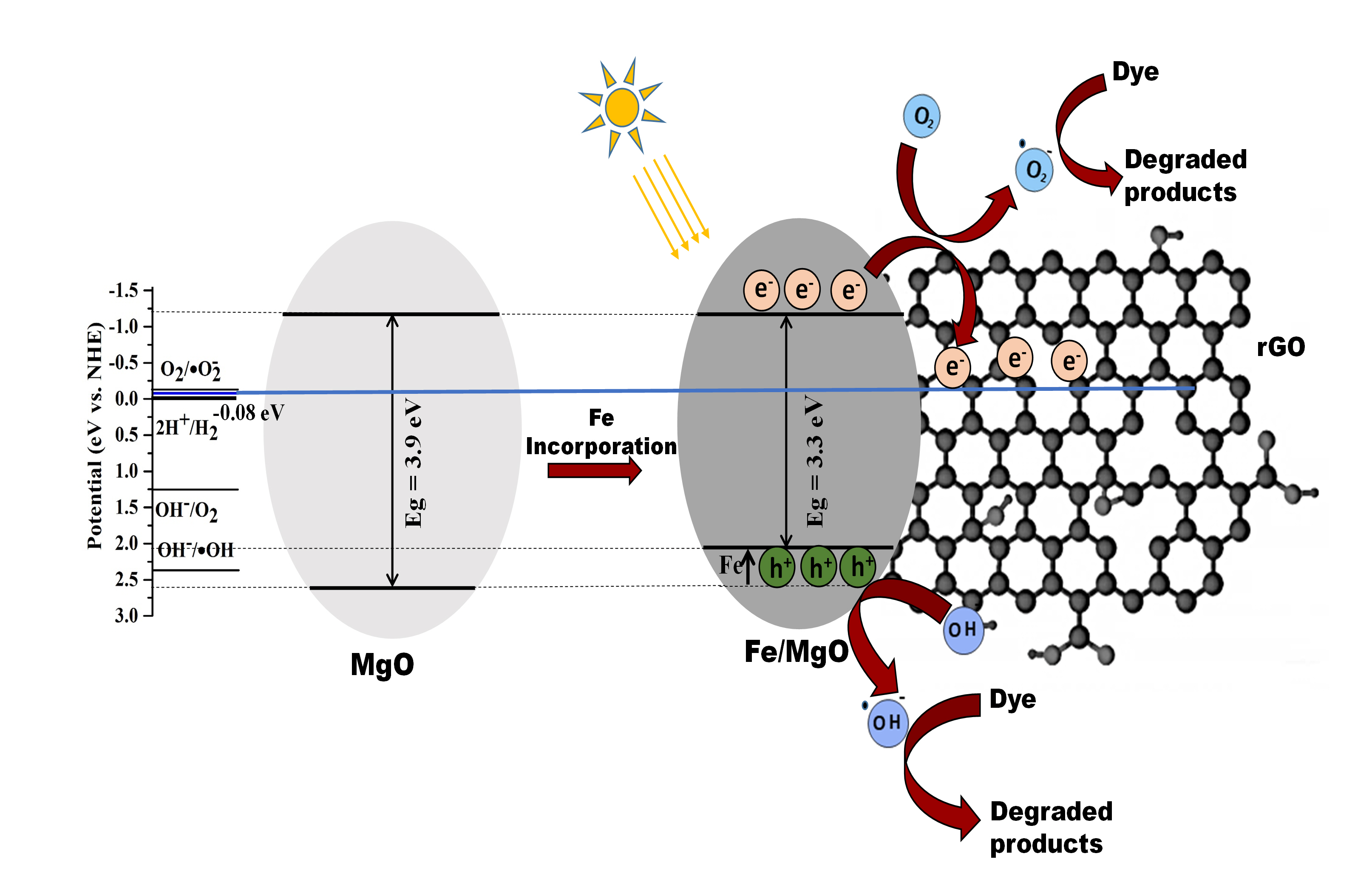}
\caption{Schematic representation of electron transfer mechanism in the photodegradation of RhB over Fe/MgO-rGO nanocomposites under solar irradiation}
\end{figure*}

A similar model can be suggested for understanding the mechanism behind the superior ability of Fe/MgO-rGO(500) nanocomposite for solar $H_{2}$  production through water splitting (ESI Fig. S7). For successful water splitting reaction to take place, CBM potential of the photocatalyst needs to be lower than the proton reduction potential (0 eV vs NHE) whereas the VBM potential needs to be greater than the oxidation potential of hydroxyl ion (1.23 eV vs NHE) \cite{Jalil}. Notably, our rGO supported nanocomposites fulfil both requirements efficiently. Therefore,
the photogenerated
holes from the nanocomposite with sufficient potential can
react with the water molecules and produce $H^{+}$ along with O\textsubscript{2}
and the photogenerated electrons evolve $H_{2}$ gas by reducing $H^{+}$.

\section{Conclusions}
We have demonstrated that incorporation of a potential electron acceptor and reservoir rGO is an excellent choice of material to maintain $e^{-}$ and   $h^{+}$ pairs separated for a long period by suppression of charge carrier recombination. This also resulted in an improved optical absorption and  reduction in band gap, all of which contributed  to enhance the photocatalytic performance of the Fe/MgO-rGO nanocomposites as compared to MgO and widely used P25 nanoparticles. Notably, the hydrogen production capability via water splitting of the synthesized nanocomposites, particularly thermally treated Fe/MgO-rGO at 500 $^{\circ}$C is significantly higher compared to that of MgO, Fe/MgO, commercial widely used P25 nanoparticles as well as other related rGO based composites. Based on the current findings, the synthesized nanocomposites can be considered as a promising candidate for numerous photocatalytic applications such as puriﬁcation of water by removing harmful pollutants, solar water disinfection, and carbon-free hydrogen production via renewable sources like water and solar energy. Our findings with profound insights into the kinetics of charge transfer may pave the way for researchers to construct and optimize new graphene-based nanocomposite photocatalysts to fulﬁll the modern technological requirements in environmental and energy-related applications.

\section*{Acknowledgements}
We sincerely acknowledge Committee for Advanced Studies and Research (CASR), Bangladesh University of Engineering and Technology for financial assistance. The Atomic Energy Center, Dhaka is also acknowledged for providing TEM facilities.
\section*{Appendix A. Supplementary data} 
 Supplementary data to this article can be found online.


\section*{Data availability}
The raw and processed data required to reproduce these findings cannot be shared at this time due to technical or time limitations.






\bibliography{PCCP} 
\bibliographystyle{rsc} 


\end{document}